\def\la{\hbox{{\lower -2.5pt\hbox{$<$}}\hskip -8pt\raise
-2.5pt\hbox{$\sim$}}}
\def\ga{\hbox{{\lower -2.5pt\hbox{$>$}}\hskip -8pt\raise
-2.5pt\hbox{$\sim$}}}
\def\ltsima{$\; \buildrel < \over \sim \;$}
\def\simlt{\lower.5ex\hbox{\ltsima}}
\def\gtsima{$\; \buildrel > \over \sim \;$}
\def\simgt{\lower.5ex\hbox{\gtsima}}

\documentstyle[epsfig]{elsart}

\begin{document}
\begin{frontmatter}
\title{Theory of synchrotron radiation: I. Coherent emission from 
ensembles of particles}
\author[lngs]{Roberto Aloisio\thanksref{corr1}} 
\address[lngs]{INFN, Laboratori Nazionali del Gran Sasso\\
SS. 17bis, Assergi (AQ), ITALY}
\thanks[corr1]{E-mail: aloisio@lngs.infn.it}
\author[Fermi]{Pasquale Blasi\thanksref{corr}} 
\address[Fermi]{INAF/Istituto Nazionale di Astrofisica\\ 
Osservatorio Astrofisico di Arcetri\\
Largo E. Fermi, 5 - 50125 Firenze, ITALY}
\thanks[corr]{E-mail: blasi@arcetri.astro.it}

\begin{abstract}
Synchrotron emission of relativistic particles in magnetic fields 
is a process of paramount importance in astrophysics. Although known
for over thirty years, there are still aspects of this radiative
process that have received little attention, mainly because they 
appear only in extreme conditions.
In the present paper, we first provide a general introduction to 
synchrotron emission, using a formalism that represents a generalization of 
the standard calculations. The use of this formalism allows us to discuss
situations in which charged particles can radiate coherently, with special 
attention for the cases in which the production occurs in the form of a
bunch of particles created in a pulse of very short duration. We calculate
the spectra of the radiation for both monoenergetic particles and 
distributions of particles with different Lorentz factors. For both cases
we study the conditions for the coherent effects to appear, and demonstrate
that in the limit of incoherent emission we reobtain the well known results. 
\end{abstract}

\end{frontmatter}

\section{Introduction}
Synchrotron radiation and its importance for astrophysics 
have been discussed in such a large number of papers 
that it is hard to believe there is anything else left to say.
The basic reviews are those in Refs. \cite{GS1,GS2} while a detailed
description of the standard theory is presented in \cite{RL}. 
Nevertheless, most previous calculations are restricted to
conditions that were considered {\it reasonable} for astrophysical
standards. These {\it reasonable} standards are now considerably 
different from those of three decades ago, when synchrotron emission
was first studied in astrophysics. 
We now know that there are situations in which 
standard calculations of synchrotron emission are not applicable. 
Two examples can be easily found and will be discussed: coherence effects 
from pulsed bunches of particles and synchrotron backreaction. 
Although some pieces of work have previously appeared in the literature,
in our opinion a complete treatment of these phenomena is still missing.
This paper will be devoted to the study of coherent synchrotron emission, 
in a very general framework, so that the conclusions may be applied to 
the cases of interest. In an accompanying paper \cite{paperII} (hereafter 
paper II) we will discuss the synchrotron backreaction, another topic 
that is rarely discussed in the literature, and for which a comprehensive 
treatment is still lacking. In paper II we will adopt the formalism introduced
here.

Coherence effects occur when there are well defined phase relations 
among the radiating particles, so that both intensity and spectra
of the resulting radiation suffer from non-negligible interference
effects. In these cases, a system of $Z$ particles with Lorentz
factor $\gamma$ has a synchrotron radiation which is up to $Z^2$ times
the spectrum of a single particle, to be compared with the incoherent
radiation, in which case the emission rate is $Z$ times the emission rate 
of a single particle. 

This is not a new point: there are many papers in which this enhancement
of the radiation was pointed out (e.g. \cite{general,saggion,benford,kirk}).
Nevertheless we think that there are important differences between these
papers and the present calculations. First, all previous papers that we are
aware of discuss the specific case of curvature radiation, mainly because the
application kept in mind is that to pulsar radio emission, where it seems
that coherence effects may be needed. Second, these past
calculations point to the evaluation of the power of the emitted radiation; 
we will devote part of this paper to point out that in case of impulsive 
coherent emission, this may be a not well defined quantity. Third, the
previous calculations take care of the coherent emission from bunches of 
particles all with the same Lorentz factor, while in the present paper we
generalize the results to the case of a spectrum of radiating particles.
As a special case, we recover the previous results.

Strictly speaking the literature that we are aware of deals with the 
process of curvature radiation, thought to be at work in pulsar magnetospheres.
This case is formally similar to the one considered here but physically 
the conditions for the occurrence of coherent effects may be quite
different.

In addition to these points, we propose a new kind of formal approach to 
the calculations of the synchrotron emission from ensemble of particles.
The new approach reproduces the results of the standard approach
but also provides new insights on the physical interpretation of those 
results. 

For the cases where coherence effects are expected, we discuss the factors 
that may be responsible for the decoherence of the emitted radiation, or, in
other words, the factors that can transform the emitted radiation from 
coherent to incoherent.

The paper is structured as follows: in section 2 we describe our formalism for
the calculation of synchrotron emission, from which the occurrence of 
coherent effects arises naturally. In section 3 we describe the concept of
radiated power in the case of coherent emission. In section 4 we use the 
approach introduced in section 2 in order to describe several features of
the coherent synchrotron emission from bunches of particles. We also 
define the condition for coherence to appear. We conclude in section 5.

\section{Synchrotron emission: the formalism}

The standard treatment of synchrotron emission from a charged particle
uses the assumption that the energy of the particle is only slightly 
affected by energy losses during a Larmor time. Within this 
assumption, which is violated in presence of backreaction 
(see papaer II), the electric field radiated by the gyrating particle 
is concentrated in a narrow beam in the direction of motion, so that 
the electric field observed by a distant observer is a short pulse, 
with duration $\sim (\gamma^3 \omega_B)^{-1}$, where $\gamma$ is the 
Lorentz factor of the particle and $\omega_B=q B/\gamma m c$ is the 
Larmor frequency.
These pulses repeat with a period $\sim 1/\omega_B$. Since the single 
pulses, for relativistic particles are very narrow, this implies that 
the frequency spectrum of the radiation, determined by the Fourier transform
of the field, is quite wide. However, most of the power is concentrated
in the high frequency range, as it is easy to understand from the short
pulses, so that the high frequency region of the spectrum is not affected
by the repetition of the pulses, but rather is dominated by the shape and 
duration of a single pulse. 
In the following we consider the two cases of a system of $Z$ particles
all with the same energy, and the case of $Z$ particles with a distribution
of energies.

\subsection{$Z$ particles with the same Lorentz factor}

A single particle would radiate energy per unit frequency and solid angle
given by \cite{jackson}:
\begin{equation}
\frac{d W}{d \omega d \Omega} = \frac{q^2 \omega^2}{4\pi^2 c}
\left | \int \hat{n} \times (\hat{n}\times\vec{\beta})
\exp \left\{ i\omega (t^\prime - \hat{n}\cdot \vec{r}(t^\prime)/c\right\}
dt^\prime \right |^2,
\label{eq:energy}
\end{equation}
where the term in the integrand is obtained through integration by parts of the
electric field, which contains $\dot{\vec{\beta}}$. Here the vector $\hat{n}$
is the unit vector in the direction of the observer (supposed here to be far 
from the emission region), $\vec{\beta}$ is the velocity vector of the particle
and $t^\prime$ is the retarded time corresponding to the radiation received
by the distant observer at time $t$.
Eq. (\ref{eq:energy}) is generalized to the case of $Z$ electrons, simply by
the substitution:
\begin{equation}
\vec{\beta} \exp\left\{ -i\omega \hat{n}\cdot \vec{r}(t^\prime)/c\right\}
\to 
\vec{\beta_i} \exp\left\{ -i\omega \hat{n}\cdot \vec{r_i}(t^\prime)/c\right\}.
\label{eq:sub}
\end{equation}
Note that in the usual treatment of synchrotron emission from an ensemble of
particles it is generally assumed that their emission is just the sum of the
emissions from the single electrons. The substitution in eq. (\ref{eq:sub})
accounts instead for the possible interference effects between electric fields
generated by different particles, labeled by the index $i$. There is another
hidden assumption in the usual treatment: not only the energies per unit 
frequency are incoherently summed, but the same procedure is adopted for
the powers. This is not what is physically more meaningful: {\bf the power
per unit frequency is proportional to the square of the Fourier transform
of the total electric field divided by the observation time}. This is not 
equivalent {\it a priori} to calculating the sum of powers and is in fact 
incorrect in some cases. As we show here, 
our approach allows for a better understanding of several points that 
do not find their explanation in the standard approach. An example
of this is provided by the study of coherence effects in the synchrotron 
emission from ensembles of particles.

If the particles have phases $\alpha_i$, we can easily write, similarly to the
case of a single particle, the following expression:
\begin{equation}
\hat{n} \times (\hat{n}\times\vec{\beta_i}) = 
-\hat{\epsilon_\perp} ~ {\rm sin}\left(\frac{v t^\prime}{a} - \alpha_i\right) +
\hat{\epsilon_\parallel} ~ {\rm cos}\left(\frac{v t^\prime}{a} - 
\alpha_i\right) {\rm sin}(\theta),
\end{equation}
where $v$ is here the modulus of the velocity, assumed to be the same for all
particles and $a$ is the Larmor radius of the gyrating particles. With these 
definitions we can also write
\begin{equation}
\frac{\hat{n}\cdot \vec{r_i}(t^\prime)}{c} = \frac{a}{c} {\rm cos}\theta {\rm sin}
\left(\frac{v t^\prime}{a} - \alpha_i\right),
\end{equation}
and the expression for the energy radiated by the ensemble of $Z$ electrons
can be written as follows:
$$
\frac{d W}{d \omega d \Omega} = \frac{q^2 \omega^2}{4\pi^2 c}
\left |\sum_{i=1}^Z \int dt^\prime \left\{-\hat{\epsilon_\perp} ~ {\rm sin}
\left(\frac{v t^\prime}{a} - \alpha_i\right) +
\hat{\epsilon_\parallel} ~ {\rm cos}\left(\frac{v t^\prime}{a} - 
\alpha_i\right)\rm{sin}\theta\right\}\right.
$$
\begin{equation}
\left.
\exp \left\{ i\omega \left[t^\prime - \frac{a}{c} \rm{cos}\theta {\rm sin}
\left(\frac{v t^\prime}{a} - \alpha_i\right)\right]\right\}\right |^2.
\label{eq:Zelec}
\end{equation}
The terms perpendicular and parallel to the direction of the magnetic 
field can always be separated so that 
\begin{equation}
\frac{d W}{d \omega d \Omega} = 
\frac{d W_\perp}{d \omega d \Omega} + \frac{d W_\parallel}{d \omega d \Omega},
\end{equation}
where the two terms can be easily derived from eq. (\ref{eq:Zelec}).
The narrow angle of the synchrotron beam implies that the observer 
can receive a signal only when $vt^\prime/a - \alpha_i \approx 0$ and
$\theta\sim 0$, so that we can adopt series expansions to write:
$$
t^\prime - \frac{a}{c} {\rm cos}\theta {\rm sin}
\left(\frac{v t^\prime}{a} - \alpha_i\right) \approx 
t^\prime -\frac{a}{c}
\left(1-\frac{1}{2}\theta^2\right)\left[\left(\frac{v t^\prime}{a}-\alpha_i
\right)-\frac{1}{6}\left(\frac{v t^\prime}{a}-\alpha_i\right)^3\right] \approx
$$
\begin{equation}
\frac{1}{2\gamma^2}\left[t^{\prime\prime}_i(1+\gamma^2\theta^2) + 
\frac{1}{3} \frac{\gamma^2 c^2 {t^{\prime\prime}_i}^3}{a^2}\right] +
\frac{a \alpha_i}{v},
\end{equation}
where we put $t^{\prime\prime}_i=t^\prime_i - a\alpha_i/v$. 
Using this expansion we obtain
$$
\frac{d W_\perp}{d \omega d \Omega} = \frac{q^2 \omega^2}{4\pi^2 c}
\left |\sum_{i=1}^Z \int dt^{\prime\prime}_i \frac{c}{a}t^{\prime\prime}_i
\right.
$$
\begin{equation}
\left.
\exp \left\{ \frac{i\omega}{2\gamma^2} \left[t^{\prime\prime}_i
(1+\gamma^2\theta^2) + \frac{1}{3}\frac{\gamma^2 c^2 
{t^{\prime\prime}_i}^3}{a^2}
\right]\right\} \exp\left\{i\omega\frac{a}{c} \alpha_i\right\}\right |^2.
\end{equation}
A similar expression can be written for $\frac{d W_\parallel}
{d \omega d \Omega}$.

We can now introduce the usual quantities, adopted in the standard literature:
\begin{equation}
\Theta_\gamma^2 = 1+\gamma^2 \theta^2, ~~~~~~y_i=\gamma 
\frac{c t^{\prime\prime}_i}
{a\Theta_\gamma},~~~~~~\eta=\frac{\omega a \Theta_\gamma^3}{3 c \gamma^3},
\end{equation}
so that changing the variable in the integral from $t^{\prime\prime}$ to
$y_i$ implies:
\begin{equation}
\frac{d W_\perp}{d \omega d \Omega} = \left(\frac{d W_\perp}
{d \omega d \Omega} 
\right)^{(1)} \left |\sum_{i=1}^Z \exp\left(i\omega\frac{a}{c}\alpha_i\right)
\right |^2
\end{equation}
and analogously
\begin{equation}
\frac{d W_\parallel}{d \omega d \Omega} = \left(\frac{d W_\parallel}
{d \omega d \Omega} 
\right)^{(1)} \left |\sum_{i=1}^Z \exp\left(i\omega\frac{a}{c}\alpha_i\right)
\right |^2,
\end{equation}

where the terms $\left(\frac{d W_\perp}{d \omega d \Omega}\right)^{(1)}$
and $\left(\frac{d W_\parallel}{d \omega d \Omega}\right)^{(1)}$ are the 
spectra expected from a single electron in the standard case. In other words,
the spectra from $Z$ particles with Lorentz factor $\gamma$ differ from the
single particle spectrum for the {\it coherence factor}
\begin{equation}
{\cal S}(Z,\omega) = 
\left |\sum_{i=1}^Z \exp\left(i\frac{\omega}{\omega_B}\alpha_i\right)
\right |^2,
\label{eq:cohfac}
\end{equation}
where we used $\omega_B=c/a$. The term ${\cal S}(Z,\omega)$ is never 
considered in standard textbook treatments because it is usually assumed that 
the particles radiate incoherently, that is to say that there is no 
interference between the electric fields radiated by different particles. 
It is instructive to study the coherence factor in some detail: we may 
rewrite it as follows:
$$
{\cal S}(Z,\omega) = \sum_{k=1}^Z \exp\left(i\frac{\omega}{\omega_B}
\alpha_k\right)
\sum_{j=1}^Z \exp\left(-i\frac{\omega}{\omega_B}\alpha_j\right) = 
$$
$$
\sum_{k=1}^Z \sum_{j=1}^Z
\exp\left[i\frac{\omega}{\omega_B}(\alpha_k-\alpha_j)\right]=
$$
\begin{equation}
\sum_{k=1}^Z \sum_{j=1}^Z 
\left\{{\rm cos}\left[\frac{\omega}{\omega_B}(\alpha_k-\alpha_j)\right]+
i~~{\rm sin}\left[\frac{\omega}{\omega_B}(\alpha_k-\alpha_j)\right]\right\}.
\end{equation}
While the second (imaginary) term trivially gives zero due to the fact
that $\alpha_k-\alpha_j$ is symmetrically distributed around zero, 
the first term is non-trivial. In the following we consider 
two cases for the coherence factor: 1) bunches of particles with the same
phase, and 2) randomly chosen phases. 

If all the phases are equal, say $\{\alpha_k\}_{k=1,Z}=0$, then the coherence
factor is trivially ${\cal S}(Z,\omega)=Z^2$ (the emission from 
Z particles radiating incoherently is simply Z times more that the 
single particle emission).

The second case of interest, in particular for astrophysics, is that of 
particles radiating incoherently. 
In this case there is no relation between the phases
$\alpha_k$ which are homogeneously (but randomly) distributed in the range
$0-2\pi$. It is useful to rewrite the coherence factor as follows 
$$
{\cal S}(Z,n) = \sum_{k=1}^Z \sum_{j=1}^Z 
{\rm cos}\left[ n \gamma (\alpha_k-\alpha_j)\right] =
Z + \sum_{k} \sum_{j\neq k} {\rm cos}\left[ n \gamma 
(\alpha_k-\alpha_j)\right] = 
$$
\begin{equation}
= Z + {\cal S^\prime}(Z,n).
\end{equation}
We studied the function ${\cal S^\prime}(Z,n)$ numerically, by generating a
large number of random configurations $\left\{\alpha_K\right\}_{k=1,Z}$ 
and deriving the probability distribution ${\cal P}({\cal S}^\prime)$ of 
the function  ${\cal S^\prime}(Z,n)$ as a function of $n$. 

Our findings can be summarized as follows: {\it a)} the function 
${\cal P}({\cal S}^\prime)$ is independent of the harmonic $n$; {\it b)}
the probability distribution can be described exceptionally well by an 
exponential function:
\begin{equation}
{\cal P}({\cal S}^\prime)\propto \exp \left\{-\frac{{\cal S}^\prime+Z}{Z}
\right\}.
\end{equation}
It is important to note that the function ${\cal P}({\cal S}^\prime)$
is peaked around ${\cal S}^\prime=-Z$, and has zero average. In other words,
while the average value of the coherence factor ${\cal S}(Z)$ is $Z$,
its most probable value is zero. This was a very unexpected result and deserves
some further comments. What happens if we have a system of radiating
particles with random phases (one configuration of them) in a magentic field?
According to our findings it seems that the emitted energy would 
likely be less than $Z$ times the power radiated by a single particle, 
and 
actually the most probable configuration is the one in which the radiated 
energy is zero.
On the other hand, if one had a large number of systems (or configurations) 
then the average energy radiated would just be the well known spectrum, equal 
to $Z$ times the one particle radiated energy. This result, though initially 
puzzling, does not imply any contradiction to known facts or observations. 
In fact, let us assume that an experiment has a frequency resolution 
$\Delta \omega$. For any reasonable choice of this resolution, there is a 
huge number of harmonics (values on $n$) in it.
One can consider a new set of random numbers 
$\left\{n\alpha_K\right\}_{k=1,Z}$ 
for different values of $n$. Each one of these configurations can be considered
as a new configuration of the system of particles with random phases, so that 
carrying out a measurement of the energy in the interval $\Delta \omega$ around
$\omega$ is equivalent to calculate the average of the coherence 
factor over many configurations, that, as found above, gives the well known 
result that the energy radiated by $Z$ particles is $Z$ times the 
energy radiated by a single particle.
We want to stress the fact however, that the reason for obtaining the
standard result is more subtle than the simple {\it incoherence} 
argument found in the literature. In principle, if one could measure the 
power at 
frequency $\omega$ with a $\Delta \omega$ less that the separation between two 
harmonics, one
would see wild fluctuations, between zero and $Z^2$ times the 
power radiated by a single particle. This result is confirmed by our numerical 
simulations, in which we calculated directly the electric field at a point
distant from the $Z$ particles and calculated the Fourier transform of it.
We detect the ``average'' value only because of our finite resolution 
instruments.

A formal demonstration of this can be provided in the following way. Let
us assume that an instrument has frequency resolution $\Delta \omega$. It
is unavoidable for $\Delta \omega$ to be much larger than $\omega_L$, the
separation between the harmonics of the synchrotron emission. On the other
hand let us choose $\Delta \omega$ so that there is negligible variation in 
the single particle emission within $\Delta \omega$.
The average spectrum measured by an experiment is then (we limit ourselves
with the perpendicular component, the parallel one being formally identical)
$$
\frac{d W_\perp}{d \omega d \Omega} = \left(\frac{d W_\perp}
{d \omega d \Omega} \right)^{(1)} \frac{1}{\Delta \omega} 
\int_\omega^{\omega+\Delta \omega} d\omega 
\left |\sum_{i=1}^Z \exp\left(i\omega\frac{a}{c}\alpha_i\right)
\right |^2 = 
$$
\begin{equation}
\left(\frac{d W_\perp}
{d \omega d \Omega} \right)^{(1)} \frac{1}{\Delta N} 
\sum_{n=N}^{N+\Delta N} \sum_{i=1}^Z
\sum_{j=1}^Z cos\left[ n \gamma (\alpha_i-\alpha_j)\right],
\end{equation}
where we used the fact that $\omega = N \omega_L$ with $N$ integer. 
The cosine term in the sum equals $\pm 1$ when $(\alpha_i-\alpha_j)=
k\pi/\gamma$, where $k$ is an integer. 
These values of $(\alpha_i-\alpha_j)$ imply a
constructive interference, while for all other values the result of the
sum over $n$ is suppressed. In the limit of many particles, we can write:
$$
\left(\frac{d W_\perp}
{d \omega d \Omega} \right)^{(1)} \frac{1}{\Delta N} 
\sum_{i=1}^Z \sum_{j=1}^Z \delta(\gamma(\alpha_i-\alpha_j)-k \pi)
= 
$$
\begin{equation}
\left(\frac{d W_\perp}
{d \omega d \Omega} \right)^{(1)} \frac{1}{\Delta N} \left[
Z \Delta N + \sum_{i\neq j}  \delta(\gamma(\alpha_i-\alpha_j)-k \pi)
\right].
\label{eq:inc}
\end{equation}
The second term in the sum defines a set of points with null measure, so that
on average
$$
\frac{d W_\perp}{d \omega d \Omega} = Z
\left(\frac{d W_\perp} {d \omega d \Omega} \right)^{(1)} 
$$
with the second term in eq. (\ref{eq:inc}) responsible for fluctuations
around the average.

Up to this point, we dealt with the calculation of the energy radiated
by an ensemble of $Z$ particles, per unit frequency and solid angle. 
What about the power (energy per unit time, per unit frequency and solid
angle)? The calculation procedure is similar to the one already described
above, but there is now an important difference in that the phases $\alpha$
contributing to the power are limited by the observation time. In the 
incoherent case, the number of particles that may contribute to the
radiation emitted during the observation time $T_{obs}$ is a fraction 
$T_{obs}/T_B(\gamma)$ of the total $Z$, where $T_B(\gamma)$ is the Larmor
rotation time of the particles. Therefore, repeating the previous steps, 
we obtain:
$$
\frac{d P_\perp}{d \omega d \Omega} =
\frac{1}{T_{obs}} \left(\frac{d W_\perp}
{d \omega d \Omega} \right)^{(1)} \frac{1}{\Delta N} 
\left[\sum_{i=1}^Z \sum_{j=1}^Z \delta(\gamma(\alpha_i-\alpha_j)-k \pi)
\right]_{0\leq \alpha\leq 2\pi T_{obs}/T_B}
= 
$$
$$
\frac{1}{T_{obs}} \left(\frac{d W_\perp}
{d \omega d \Omega} \right)^{(1)} \frac{1}{\Delta N} \left[
Z \Delta N  \frac{T_{obs}}{T_B} + 
\left\{\sum_{i\neq j} \delta(\gamma(\alpha_i-\alpha_j)-k \pi)
\right\}_{0\leq \alpha\leq 2\pi T_{obs}/T_B}\right]\approx
$$
\begin{equation}
\left(\frac{d W_\perp}{d \omega d \Omega} \right)^{(1)} \frac{Z}{T_B},
\end{equation}
where in the last step we neglected the fluctuation term. This is the
well known result of particles radiating incoherently.

\subsection{$Z$ particles with a distribution of Lorentz factors}

We consider here the case of $Z$ particles having a spectrum of energies
and some arbitrary phase distribution. We start from the same basic expression
found in the previous section:
\begin{equation}
\frac{d W}{d \omega d \Omega} = \frac{q^2 \omega^2}{4\pi^2 c}
\left | \sum_{i=1}^Z \int \hat{n} \times (\hat{n}\times\vec{\beta_i})
\exp \left\{ i\omega (t^\prime - \hat{n}\cdot \vec{r_i}(t^\prime)/c\right\}
dt^\prime \right |^2,
\label{eq:energyZ}
\end{equation}
where $\vec{\beta_i}=v_i/c$ is the dimensionless velocity vector of the $i$-th
particle, calculated at the retarded time $t^\prime$. Following the same steps
as in section 2, we obtain the following exrpessions for the perpendicular and
parallel components of the emitted spectra:
\begin{equation}
\frac{d W_\perp}{d \omega d \Omega} = \frac{q^2 \omega^2}{4\pi^2 c}
\left | \sum_{i=1}^Z \int dt^\prime {\rm sin}\left( \frac{v_i t^\prime}{a_i}-
\alpha_i\right) \exp\left\{ i \omega\left[ t^\prime -\frac{a_i}{c}{\rm cos} 
\theta
{\rm sin}\left(\frac{v_i t^\prime}{a_i} - \alpha_i\right)\right]\right\}
\right |^2,
\end{equation}

\begin{equation}
\frac{d W_\parallel}{d \omega d \Omega} = \frac{q^2 \omega^2}{4\pi^2 c}
\left | \sum_{i=1}^Z \int dt^\prime {\rm cos}\left( \frac{v_i t^\prime}{a_i}-
\alpha_i\right) {\rm sin}\theta 
\exp\left\{ i \omega\left[ t^\prime -\frac{a_i}{c}{\rm cos} \theta
{\rm sin}\left(\frac{v_i t^\prime}{a_i} - \alpha_i\right)\right]\right\}
\right |^2.
\end{equation}
Even in this case, the bursts of radiation reach the observer only when 
$v_i t^\prime/a_i\approx a_i$ and $\theta\sim 0$, so that we can still use
the expansions adopted in the previous section and obtain
$$
\frac{d W_\perp}{d \omega d \Omega} = \frac{q^2 \omega^2}{4\pi^2 c}
\left | \sum_{i=1}^Z \int \left(\frac{a_i \Theta_{\gamma_i}}{\gamma_i c}
\right) dy_i \frac{\Theta_{\gamma_i}}{\gamma_i} y_i 
\right.
$$
\begin{equation}
\left.
\exp\left\{\frac{i \omega}
{2 \gamma_i^2}\left[\frac{a_i \Theta_{\gamma_i}^3}{c \gamma_i} y_i +
\frac{1}{3} \frac{a_i \Theta_{\gamma_i}^3}{\gamma_i c} y_i^3\right]\right\}
\exp\left\{ i \omega \frac{a_i}{c} \alpha_i\right\}\right |^2,
\end{equation}
where
\begin{equation}
\Theta_{\gamma_i}^2 = 1+\gamma_i^2 \theta^2, ~~~~~~y_i=\gamma_i 
\frac{c t^{\prime\prime}_i}{a_i\Theta_{\gamma_i}},~~~~~~\eta_i=
\frac{\omega a_i \Theta_{\gamma_i}^3}{3 c \gamma_i^3},
\end{equation}
A similar formula holds for the parallel component 
$\frac{d W_\parallel}{d \omega d \Omega}$.
After some algebraic steps, we can rewrite the spectrum in the following form:
\begin{equation}
\frac{d W_\perp}{d \omega d \Omega} = \frac{q^2 \omega^2}{3\pi^2 c}
\sum_{i=1}^Z \sum_{j=1}^Z \frac{a_i a_j \Theta{\gamma_i}^2 \Theta_{\gamma_j}^2}
{c^2 \gamma_i^2 \gamma_j^2} K_{2/3}(\eta_i) K_{2/3}(\eta_j) \exp \left\{
i\frac{\omega}{c} (a_i \alpha_i - a_j \alpha_j)\right\}
\label{eq:Wperp}
\end{equation}
and similarly:
\begin{equation}
\frac{d W_\parallel}{d \omega d \Omega} = \frac{q^2 \omega^2}{3\pi^2 c}
\sum_{i=1}^Z \sum_{j=1}^Z \theta^2 
\frac{a_i a_j \Theta{\gamma_i} \Theta_{\gamma_j}}
{c^2 \gamma_i \gamma_j} K_{1/3}(\eta_i) K_{1/3}(\eta_j) \exp \left\{
i\frac{\omega}{c} (a_i \alpha_i - a_j \alpha_j)\right\},
\label{eq:Wpar}
\end{equation}
where $K_{2/3}$ and $K_{1/3}$ are modified Bessel functions.

In this case, defining a coherence factor is not as simple as for the 
case of $Z$ particles with the same Lorentz factor. Nonetheless,
it is possible to consider some cases of interest. In the case of particles
with no phase relation, the expressions in eqs. (\ref{eq:Wperp}) and 
(\ref{eq:Wpar}) allow us to recover the standard result. Let us consider
again a narrow frequency range $\Delta \omega$ containing many harmonics, 
but still small enough that the non-oscillatory term in eqs. (\ref{eq:Wperp}) 
and (\ref{eq:Wpar}) can be taken as constant in $\omega$. 
Therefore the average spectrum in the frequency range $\Delta \omega$ 
around the frequency $\omega=N \omega_L$ can be written as:
$$
\frac{d W_\perp}{d \omega d \Omega} = \frac{q^2 \omega^2}{3\pi^2 c}
\sum_{i=1}^Z \sum_{j=1}^Z \frac{a_i a_j \Theta{\gamma_i}^2 \Theta_{\gamma_j}^2}
{c^2 \gamma_i^2 \gamma_j^2} K_{2/3}(\eta_i) K_{2/3}(\eta_j) 
\frac{1}{\Delta N} \sum_{n=N}^{N+\Delta N} {\rm cos}\left[ 
n (\gamma_i \alpha_i - \gamma_j \alpha_j)\right] =
$$
\begin{equation}
 \frac{q^2 \omega^2}{3\pi^2 c}
\sum_{i=1}^Z \sum_{j=1}^Z \frac{a_i a_j \Theta{\gamma_i}^2 \Theta_{\gamma_j}^2}
{c^2 \gamma_i^2 \gamma_j^2} K_{2/3}(\eta_i) K_{2/3}(\eta_j) 
\delta_{ij}
\end{equation}
where we eliminated the fluctuation term on the same basis used in the 
previous section for the case of $Z$ electrons with the same Lorentz factor.
At this point it is easy to reobtain the well known result, simply by 
passing from the sums to the integrals:
\begin{equation}
\frac{d W_\perp}{d \omega d \Omega} = \frac{q^2 \omega^2}{3\pi^2 c}
\int d\gamma N(\gamma)
\frac{a(\gamma)^2 \Theta_{\gamma}^4}{c^2 \gamma^4} K_{2/3}^2(\eta). 
\end{equation}
As usual, a similar expression is obtained for the parallel component.

Also in this case, there is a difference between the energy spectrum and
the power. Following the guidelines adopted in the previous section, one
easily obtains for the power the following expression:
\begin{equation}
\frac{d P_\perp}{d \omega d \Omega} = \frac{q^2 \omega^2}{3\pi^2 c}
\int d\gamma N(\gamma) \frac{1}{T_B(\gamma)} \frac{a(\gamma)^2 
\Theta_{\gamma}^4}{c^2 \gamma^4} K_{2/3}^2(\eta). 
\label{eq:Power_perp}
\end{equation}
We stress again that this procedure is correct only if the effect of 
backreaction is negligible (paper II). 

In the case of particles all with the same phase, the exponential term in 
eqs. (\ref{eq:Wperp}) and (\ref{eq:Wpar}) becomes unity, so that we can 
write, after passing to the continuum in $\gamma$, the following expressions:

\begin{equation}
\frac{d W_\perp}{d \omega d \Omega} = \frac{q^2 \omega^2}{3\pi^2 c}
\left[ \int_{\gamma_{min}}^{\gamma_{max}} d\gamma N(\gamma)
\frac{a(\gamma) \Theta_{\gamma}^2}{c \gamma^2} K_{2/3}(\eta(\gamma))\right]^2,
\label{eq:zerophases}
\end{equation}
and
\begin{equation}
\frac{d W_\parallel}{d \omega d \Omega} = \frac{q^2 \omega^2}{3\pi^2 c} 
\theta^2 \left[ \int_{\gamma_{min}}^{\gamma_{max}} d\gamma N(\gamma)
\frac{a(\gamma) \Theta_{\gamma}}{c \gamma} K_{1/3}(\eta(\gamma))\right]^2.
\label{eq:zerophasespar}
\end{equation}
Note that in the special case $N(\gamma)=Z\delta(\gamma-\gamma_0)$, we
reobtain the result found in the previous section for the case of Z particles
with the same Lorentz factor $\gamma_0$, which is amplified by a factor 
$Z^2$ with respect to the case of a single particle.

In the case of a power law spectrum of particles all with the same phase 
some useful results can be found by using the asymptotic expressions for 
the modified Bessel functions (it is useful to notice 
the difference between the result in eqs. (\ref{eq:Wperp}) and (\ref{eq:Wpar})
and the corresponding expressions for the standard case of particles 
radiating incoherently, where the Bessel functions appear squared).

We use the following approximation for the modified Bessel functions:
$$ 
K_\nu (x)=\frac{2^{\nu}\pi}{3^{1/2}}\frac{1}{\Gamma(1-\nu)} x^{-\nu}
~~~\rm{if}~~~ 0\leq x\leq 1~~~\rm{and}~~~ 0~~~ \rm{otherwise},
$$
which is accurate enough to allow us to derive the correct spectral slopes
and magnitudes. Using the definition $a(\gamma)=m c^2 \gamma/q B$ and 
the above approximation for the $K_\nu$, we obtain for the perpendicular
component:
\begin{equation}
\frac{d W_\perp}{d \omega d \Omega} = \frac{m^2 c \omega^2}{3\pi^2 B^2}
\left[ \int_{\gamma_{min}(\theta,\omega)}^{\gamma_{max}} d\gamma 
\frac{N(\gamma)}{\gamma} \Theta_\gamma^2 
\frac{2^{2/3}\pi}{3^{1/2}}\frac{1}{\Gamma(1/3)} 
\left(\frac{m c \omega}{3 q B \gamma^2}\Theta_\gamma^3\right)^{-2/3}\right]^2,
\end{equation}
while similar expressions can be derived for the parallel component.
The lower limit in the integral is calculated by requiring that
the expansion for the Bessel function is valid, which is to say:
\begin{equation}
\frac{m c \omega}{3 q B \gamma^2} \left[ 1 + \gamma^2 \theta^2 \right]^{3/2}
\leq 1.
\end{equation}
Since most of the synchrotron emission comes from the region $\theta\gamma<1$,
we can expand the term $\left[ 1 + \gamma^2 \theta^2 \right]^{3/2}$ around
$\theta\gamma\sim 0$ (even at $\theta\gamma\sim 1$ the expansion gives about
$\sim 10\%$ accuracy, retaining up to the second term in the expansion). This 
implies:
\begin{equation}
\gamma_{min}(\theta,\omega) = \left[\frac{3 q B}{m c \omega} - 
\frac{3}{2}\theta^2 \right]^{-1/2}.
\end{equation}
Therefore we obtain:
\begin{equation}
\frac{d W_\perp}{d \omega d \Omega} = 
\frac{(6 q)^{4/3}}{9 c^{1/3} \Gamma^2(1/3)} (m \omega)^{2/3} B^{-2/3}
\left[\frac{N_0}{p-\frac{4}{3}}\right]^2 \left[\frac{3 q B}{m c \omega} - 
\frac{3}{2}\theta^2 \right]^{p-\frac{4}{3}}
\end{equation}
where we assumed that that the spectrum of particles is $N(\gamma)=N_0
\gamma^{-p}$ with $p>4/3$, which usually reflects the behaviour of
astrophysical sources.
In order to obtain the final result, we only need to integrate over the 
solid angle $d\Omega = 2\pi \rm{sin}(\alpha_p) d\theta$, where $\alpha_p$
is the pitch angle. In the following we assume that $\rm{sin}(\alpha_p)=1$
for simplicity, which does not affect the generality of our calculations.
The integration over $\theta$ immediately provides the frequency spectrum
that can be shown to be $\propto \omega^{-p+3/2}$. 
As expected, the spectrum of the coherent radiation scales as the square
of the normalization constant $N_0$. 

\section{Power per unit frequency versus Energy per unit frequency}

When the coherence effects are taken into account the distinction between 
the concepts of power and energy per unit frequency need to be carefully
analyzed. In fact the differentiation between these two concepts becomes
mandatory in the case of coherent emission from bunches of particles with
a spectrum of energies.

Let us start with the incoherent case for an ensemble of particles with
energy spectrum $N(\gamma)$: the power is then given by the 
well known expression, eq. (\ref{eq:Power_perp}) (and the corresponding
term for the parallel component). This expression is motivated by the 
assumption that the emission can be averaged over a Larmor time. This makes 
a term $1/\gamma$ appear within the integral over the spectrum. 
If the backreaction effect is negligible (that is if the particles do not
lose appreciable energy during the Larmor gyration time), this assumption
is well justified, and eq. (\ref{eq:Power_perp}) provides the power averaged
over a Larmor time of each particle.

Let us now consider the case of an ensemble of particles that are in
perfect phase with each other, so that the emitted radiation is coherent.
In this case the procedure used above for the definition of the power is 
not correct. It is clear in fact that, if the particles with different Lorentz
factor do radiate coherently for some time interval $\Delta t$, this interval 
is likely to be immediately after the production of these particles, while 
after a short while the different Lorentz factors will easily break the 
coherence due to different Larmor radii of the particles. In this case, the
power is not determined by an average over the Larmor times of the particles,
but rather by the occurrence of the processes responsible for the production
of compact bunches of particles. Even formally it would then be incorrect
to simply divide by the gyration time in the integrals in eqs. 
(\ref{eq:zerophases}) and (\ref{eq:zerophasespar}). 

In the previous calculations this crucial problem of distinguishing 
power and energy per unit frequency was not considered and the particles
were assumed to move periodically along a circle. This assumption was 
adopted because the interest was focused on the curvature radiation of
charged particles moving along curved magnetic field lines, that played
therefore the role of tracks over which the particles were forced to move. 
This is not the case when synchrotron emission is considered, and even more
when the radiating particles are not monoenergetic.

\section{Coherence and decoherence}

In the previous sections we have discussed several idealized situations in 
which the coherent synchrotron radiation appears. However in more realistic
situations it is necessary to identify the conditions in which the emission
can really be considered coherent and is eventually kept as such in time. 
Moreover, the calculations carried out in the previous section concentrate 
on the energy emitted by the particles, rather than the power. In the case 
of coherent emission the concept of radiated power needs special care. 

The coherence factor introduced in eq. (\ref{eq:cohfac}) contains the
main physical ingredients for the study of the coherence effects in 
ensembles of particles with the same energy. In the case of $Z$ particles 
with null phase difference, it is straightforward to obtain that the
power from the ensemble of particles is $Z^2$ times larger than the power
radiated by a single particle. Let us study the more interesting and 
realistic case of a bunch of particles whose spread in phase $\Delta\alpha$ 
is small but finite. 

We distinguish three regimes: 1) $n\gamma\Delta\alpha\ll 1$; 
2) $n\gamma\Delta\alpha\sim 1$; 3) $n\gamma\Delta\alpha\gg 1$. 

The starting point is again eq. (\ref{eq:cohfac}):
$$
{\cal S}(Z)=\sum_{i=1}^Z \sum_{j=1}^Z {\rm cos}\left[ n \gamma 
(\alpha_i-\alpha_j)\right].
$$
In case (1), $n\gamma (\alpha_i-\alpha_j)\leq n\gamma\Delta\alpha\ll 1$,
therefore the cosine terms can be expanded in series and neglecting terms
of higher orders, we simply obtain the coherent result ${\cal S}(Z)=Z^2$.

In the opposite limit (case 3) $n\gamma\Delta\alpha\gg 1$, the 
rapid oscillations of the cosine function are averaged out within a
sufficiently large frequency range (or equivalently on a large number 
of configurations of the system), so that the standard incoherent result 
is recovered. 

In the intermediate range, $n\gamma\Delta\alpha\sim 1$, we can carry out
the passage from the sum to the integral:
$$
{\cal S}(Z)=\sum_{i=1}^Z \sum_{j=1}^Z {\rm cos}\left[ n \gamma 
(\alpha_i-\alpha_j)\right] = 
\int_{-\Delta\alpha/2}^{\Delta \alpha/2} d\alpha 
\int_{-\Delta\alpha/2}^{\Delta \alpha/2} d\beta
\left(\frac{Z}{\Delta\alpha}\right)^2 {\rm cos}\left[ n \gamma 
(\alpha-\beta)\right] = 
$$
\begin{equation}
= \left(\frac{Z}{\Delta\alpha}\right)^2 
\frac{4 \sin^2 \left[\frac{n \gamma \Delta\alpha}{2}\right]}{n^2 \gamma^2}.
\end{equation}
Note that this equation recovers the result ${\cal S}(Z)=Z^2$ obtained for 
particles with null phase difference (or $\Delta\alpha\ll 1/n\gamma$).
When $n\gamma\Delta\alpha\sim 1$, the spectra should be characterized by the 
peculiar peaks with decreasing height of the function $\sin^2(x)/x^2$.

Our calculations of the spectrum of the radiation emitted by a bunch of
monoenergetic particles with different phase spreads $\Delta\alpha$ is
plotted in fig. 1 (in particular we plot the perpendicular component
with $\theta=0$). We parametrize the spread as 
\begin{equation}
\Delta\alpha=\frac{\xi}{\gamma^3},
\end{equation}
so that the coherence condition reads now $n\ll \gamma^2/\xi$. In other
words, for fixed $\gamma$ and $\xi$ (or $\Delta\alpha$) the coherence
occurs at low freqeuncies. The curves plotted in fig. 1 refer, as indicated,
to $\xi=10^6$ (fully incoherent case), $\xi=10^4$, $\xi=10$, $\xi=1$
and $\xi=10^{-2}$. 

For the cases $\xi=10^4$ and $\xi=10$ the transition frequency sits
in a place that allows us to see the transition from coherence to 
incoherence, in the form of the typical oscillations of the function
$sin^2 x/x^2$, as explained above. Note also that in the coherence regions
the ratio of the coherent to incoherent spectra is exactly equal to 
$Z$, the number of particles, taken here to be $Z=1000$. One can easily
check that the slope of the power law pieces are those expected for the 
one particle spectra per unit solid angle at $\theta=0$.
 
\begin{figure}[thb]
 \begin{center}
  \mbox{\epsfig{file=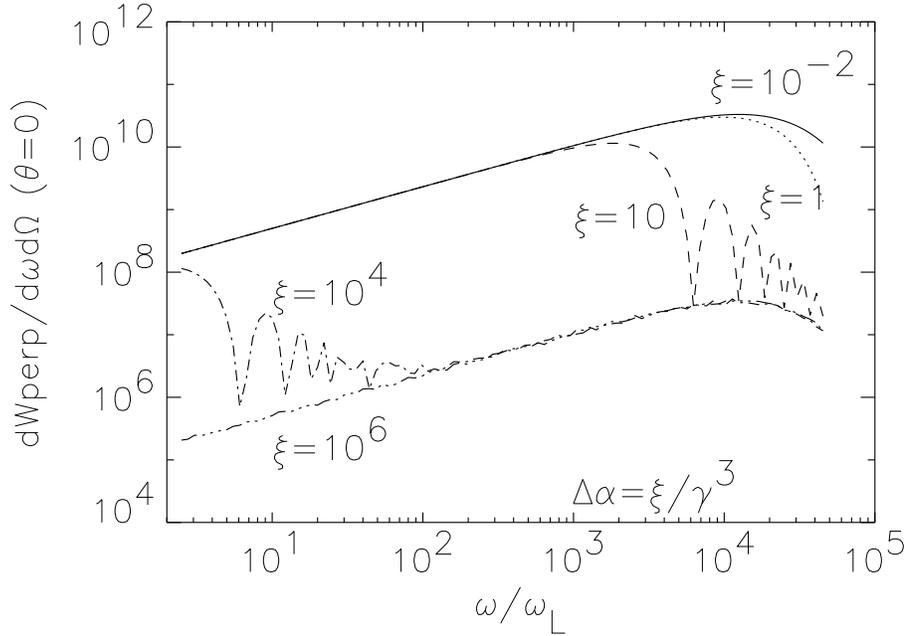,width=13.cm}}
  \caption{\em {Energy emitted per unit frequency in the perpendicular
component for $\theta=0$ for $\xi=10^{-2},~1,~10,~10^4,~10^6$ in case of
a monoenergetic bunch of particles. The coherence effect at low frequencies 
is evident.
}}
 \end{center}
\end{figure}

The case of a bunch of particles with spread $\Delta\alpha$ and with 
a distribution of Lorentz factors is more difficult but still treatable. 
We can still define a phase spread as 
\begin{equation}
\Delta\alpha=\frac{\xi}{\gamma_{max}^3}
\end{equation}
and explore the range of values of $\xi$ and the frequency $n$
for which we have coherence. 
It is then easy to show that $n(\alpha_i\gamma_i-\alpha_j\gamma_j)\leq 
n \gamma_{max} \Delta\alpha$; therefore, if $n \gamma_{max}\Delta\alpha\ll 1$,
the arguments of the cosines in eq. (\ref{eq:Wperp}) can be expanded in 
series and eq. (\ref{eq:Wperp}) gives eq. (\ref{eq:zerophases})
(similar calculations can be carried out for the parallel component), already 
obtained for the case of null phase difference among the particles. 
Similarly to the monoenergetic case the frequency at which the transition 
from the coherent to the incoherent regime occurs is 
$n\sim \gamma_{max}^2/\xi$. This is readily visible in fig. 2, where
we plotted our calculations for $\gamma_{min}=5$, $\gamma_{max}=1000$,
$Z=1000$ and the values of $\xi$ indicated in the figure.  
 
\begin{figure}[thb]
 \begin{center}
  \mbox{\epsfig{file=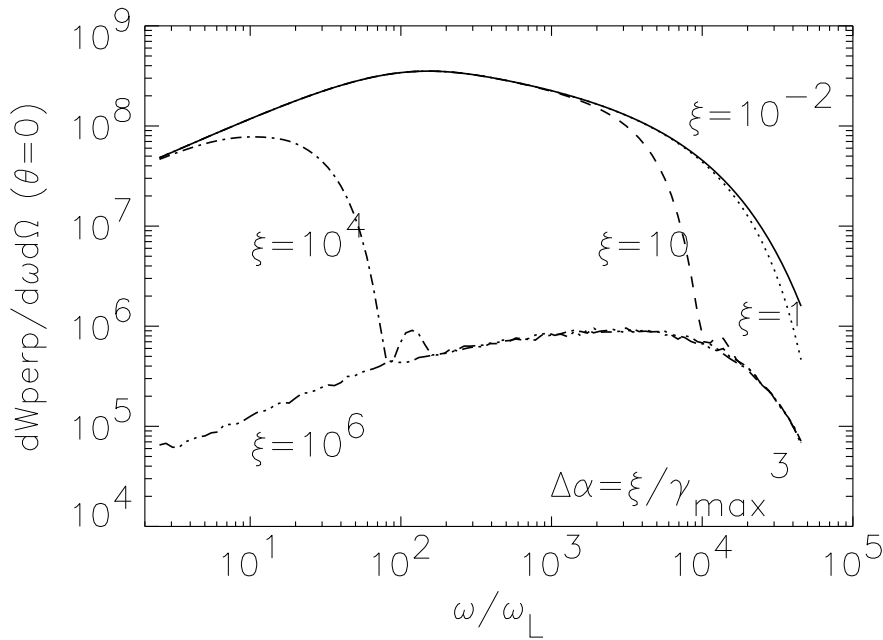,width=13.cm}}
  \caption{\em {Eenrgy emitted per unit frequency in the perpendicular
component for $\theta=0$ for $\xi=10^{-2},~1,~10,~10^4,~10^6$ for a power
law spectrum of particles with $p=2.1$.
}}
 \end{center}
\end{figure}

A bunch of particles radiating coherently can be produced by plasma 
effects or due to the intrinsic properties of the system, for instance
the physical beaming in a pulsed source. 

Once a bunch of particles radiating coherently has been generated
it is important to study the conditions under which it keeps on radiating
coherently. This point is actually related to the issue of calculating
the power versus the energy radiated coherently. 
The simplest case is again that of a monoenergetic bunch. In this case,
if the magnetic field were perfectly homogeneous over scales comparable with 
the Larmor radius of the gyrating particles, the coherence would be preserved.
Any inhomogeneity in the magnetic field may imply a phase shift among 
the particles without affecting the coherent radiation process, provided 
such shift satisfies the condition written above. 

The case of a bunch of particles with different Lorentz factors is 
complicated by the fact that the coherence is distroyed by the very
motion of the particles with different Lorentz factors. This decoherence
effect is time dependent and implies a progressive phase shift among 
the particles. In addition to this effect, there is the decoherence 
action due to possible inhomogeneity in the magnetic field. However,
in this case, the situation is also affected by the different Larmor
radii of particles with different Lorentz factors, that clearly favors
the decoherence spread of the bunch. It is therefore very difficult for
the coherence effects from a non-monoenergetic bunch to be kept in 
time. 

\section{Conclusions}

We studied the theory of coherent synchrotron emission in the perspective of
possible astrophysical applications. The theory is obtained from the general
treatment of radiation from an ensemble of particles, so that the usual 
results are easily recovered when the coherence effects are not relevant.

Our conclusions can be summarized as follows:

{\it i)} An ensemble of $Z$ monoergetic particles in perfect phase 
radiate coherently at any frequency, and the total radiated energy 
is $Z^2$ larger than the energy radiated by a single particle at the 
same frequency. This can be easily interpreted by recalling that the
synchrotron spectra are proportional to the square of the electric charge
of a particle: coherent emission from $Z$ particles can be thought as
the synchrotron emission of a single particle with charge $Z$.

{\it ii)} If $Z$ monoenergetic particles have a phase spread $\Delta\alpha$,
the coherence is limited to frequencies $n\ll 1/\Delta\alpha\gamma$. At 
higher frequencies the incoherent result is gradually recovered.

{\it iii)} In the case of $Z$ particles with a spectrum $N(\gamma)\propto
\gamma^{-p}$ in the range $\gamma_{min}\leq\gamma\leq\gamma_{max}$ and 
spread in phase over an angle parametrized as $\Delta\alpha=\xi/
\gamma_{max}^3$, the condition for coherence is that $n\ll \gamma_{max}^2/\xi$.
At high frequencies, $n\gg \gamma_{max}^2/\xi$, the incoherent result is
recovered.

{\it iv)} The coherent emission from a bunch of particles is not likely
to be stable in time: in the case of a monoenergetic bunch, small 
inhomogeneities in the magnetic field structure introduce phase shifts
among the particles, so that the coherence condition may be easily broken.
In the case of a bunch of particles with a distribution of Lorentz factors,
inhomogeneity of the magnetic field and the fact that the Larmor radii
of the particles are different makes the stability of the coherence very
problematic. In both cases the coherent emission can however be generated
at the time of formation of the bunch, for instance in pulsed events.

Despite the difficulty in maintaining the coherent character of the radiation,
there seem to be several situations in which invoking the coherence appears
to provide the most reasonable explanation for the observations. One case is
that of the radio emission from pulsars and is discussed in \cite{michel}.
In this case the radiation is however most likely curvature radiation
rather than synchrotron emission. Although the two mechanisms are very 
similar, there are also technical differences between the two. The second case
of possible coherent emission is related to the radio brightness of 
jets in active galactic nuclei, when the brightness temperature is 
$T_B>3\times 10^{17}$ K. This case has been mentioned in the literature
(e.g. \cite{spada,ghisellini}) but never treated in detail.

{\bf aknowledgments} We are very grateful to F. Pacini and A. Olinto for 
several constructive discussions and to M. Salvati and T. Stanev for a 
critical reading of the manuscript. We are also grateful to the anonymous 
referee for the interesting remarks on the paper.

\end{document}